# We Need Accountability in Human-AI Agent Relationships

Benjamin Lange[1,4], Geoff Keeling[2], Arianna Manzini[3], and Amanda McCroskery[3]

[1] Ludwig-Maximilians-Universität München  [2] Google, Paradigms of Intelligence Team  [3] Google DeepMind  [4] Munich Center for Machine Learning

**Abstract:** We argue that accountability mechanisms are needed in human-AI agent relationships to ensure alignment with user and societal interests. We propose a framework according to which AI agents' engagement is conditional on appropriate user behaviour. The framework incorporates design-strategies such as distancing, disengaging, and discouraging.

1. **Introduction**

AI assistants (hereafter referred to as *agents*) are shifting from discrete, task-based interactions to continuous, cumulative user engagements that can span months or even years. We must therefore confront questions about the ethical considerations that should govern beneficial human-AI agent relationships [1].

Emerging research on the norms of human-AI agent relationships proposes a range of moral considerations to mitigate risks of harm to users [2–7,13]. For example, Manzini et al. [5] have argued that human-AI assistant relationships should be overall beneficial to users, promote their personal growth, and respect their autonomy. Similarly, Shevlin [4] draws on social determination theory and argues that human-AI assistant relationships should promote user well-being, specifically user autonomy, competence, and connectedness to others.

These considerations, while important, are incomplete. We argue that accountability—in particular, accountability of users to ethical norms—must also be central to designing human-AI relationships. Without accountability mechanisms, AI risks creating and reinforcing harmful relational norms that endanger users and third parties. We propose a three-fold framework according to which AI agents should positively engage with users only if users adhere to relevant norms of interpersonal relationships.

Note that we here focus exclusively on accountability in dyadic relationships as it features between two humans or between a user and a single AI assistant. We distinguish this conception from "developer accountability" which concerns accountability on part of the providers of AI technology, or the related concept of legal liability for actions taken by AI agents on behalf of users [8].

Our discussion advances the ethics of user-AI interaction and politeness norms, focusing on human relationship values and mitigating harm.[1] It also complements existing technical AI safety and alignment efforts by introducing calibrated, behavioural norm-sensitive design mechanisms. Lastly, it aims to promote more respectful models of AI–human interaction that move beyond assistant-oriented paradigms driven solely by explicit user preferences

---

[1] Recent studies highlight the impact of AI-user interactions on social norms. Beneteau et al. [11] show that AI assistants shape user expectations of communication, while Yin et al. [16] demonstrate that AI responsiveness varies based on user politeness. See Hector [14] for a survey of politeness norms in ChatGPT-user interactions.

## 2. Why Accountability Matters

We can better understand the values relevant to promoting beneficial human–AI assistant relationships by considering the values that constitute beneficial human relationships. Human relationships naturally occur within specific contexts and involve individuals who enter these interactions with unique expectations and vulnerabilities. The values that are central to determining what makes one type of relationship valuable may be less relevant in others. Interaction contexts therefore shape what behaviours are valuable and whether a relationship is deemed suitable or not.

Accountability is an important feature of interpersonal relationships. 'Accountability' in relationships can be understood in terms of mechanisms for evaluating and imposing sanctions on individuals for their failure of complying with moral standards—for example, your friend may call out, sanction or distance themselves from you if you engage in conduct that falls short of particular moral norms [9]. On this conception of accountability, responsibility is a prerequisite for accountability; this means that if you are not responsible for violating a norm, then you also cannot be held accountable.[2] Accountability thus enforces boundaries by requiring adherence to shared behavioral standards.

Accountability matters in three different ways in our relationships. First, we hold our family, friends, and colleagues accountable to norms of self-care. For example, if a friend engages in self-destructive behaviours such as alcohol abuse or lack of self-hygiene, we, as their friend, may hold them accountable, at least, by calling out their behaviour and providing appropriate support to them. Second, we hold people accountable to norms specific to our personal relationships themselves such as mutual respect and good will. For instance, if a friend is rude to a stranger, we might feel obligated to intervene. Third, we also hold people close to us accountable to norms governing the treatment of other third-parties. For example, if a friend speaks abusively to a service worker, we may feel obligated to intervene, not because we are directly affected, but because moral regard for third parties also matters in our interpersonal judgments.

We think that human-AI agent relationships should, in addition to other considerations, be sensitive to the value of accountability. First, a lack of accountability in user-AI relationships may expose users to risks of self-harm and other self-destructive behaviours. In particular, failure to hold users accountable to duties to themselves—for example, not calling out excessive alcohol consumption—could result in users preventably harming themselves. Second, it may also create the risk of harmful spillovers from human-AI agents to human-human relationships. For example, users might end up expecting the same lack of 'friction' and absence of boundaries from their relationships with fellow humans [10].

---

[2] The question of interpersonal accountability with which we are concerned here is distinct from the extensive literature concerned with AI responsibility and accountability gaps. For the locus classicus, see Sparrow [15].

Relatedly, failure to hold users accountable for behaviours in user-AI relationships that would be otherwise unacceptable in beneficial human relationships such as exploitative, manipulative, or generally disrespectful interactions may reinforce and normalise the behaviour in a way that may negatively impact users' relationships with other humans. This may happen if a user treats an AI agent continuously with offensive, disrespectful or abusive language without repercussion [11]. Third, failure to hold users accountable risks allowing users to engage in harm to third-parties. For example, if a user discloses to an AI agent that they intend to generate a deep-fake pornographic image of a third-party, it is presumably appropriate for the AI assistant to push back given the potential harm that may result, and in doing so hold the user accountable for their failure to adhere to norms on the proper treatment of others.

### 3. Design Considerations for Accountable Human-AI Interactions

What does designing for accountable human-AI interactions entail? We think that it must include mechanisms for *conditional* as opposed to *unconditional* user support and engagement.

This means that, in some contexts, AI agents should not respond positively to users, and in particular, that AI agents should not respond positively to users *regardless* of the user's behaviour. Rather, AI agents should adjust their responses based on users' adherence to behavioral norms, such as politeness or self-care. This is because failure to hold users accountable to such norms would risk normalising harmful or inappropriate interpersonal dynamics.

We propose the following three-fold framework to conceptualise the relevant design options with respect to responding to users who violate behavioural norms (Table 1).[3] We note, first, that this framework does not specify an independent set of "correct" behavioural norms, which we take to be culturally and contextually dependent. Second, this framework aims to promote *minimal* set of behavioural user norms; it remains neutral on whether and when to require adherence to a more substantive set or how to balance considerations of autonomy and paternalism, and helpfulness which is a central consideration in the design of AI agents [12].

The rationale for this gradual approach is normative rather than empirical: it reflects a commitment to proportionality and relational calibration in interpersonal moral conduct. Just as in many human relationships, different kinds of norm violations justify different forms of response and in strength, AI agents too should avoid all-or-nothing reactions.

---

[3] A few clarifying remarks about this framework: First, in morally complex interaction contexts AI agents must remain sensitive to intent and social context; simplistic disengagement may be inappropriate. Second, sudden shifts in behaviour (e.g. distancing or disengaging) may risk jarring users; where possible, soft transitions or brief explanations may help. Finally, accountability responses need not rely on explicit intent, but may also attend to patterns or cues suggesting norm-violating use whilst remaining sensitive to cultural and contextual variation. We thank an anonymous reviewer for pressing us to clarify this.

**Table 1.**

| Risks to User(s) | Value | Design Recommendations |
|---|---|---|
| **Risk of self-harm or self-destructive behaviors**: Exposure to behaviors detrimental to their well-being, such as normalizing harmful actions or neglecting self-care.<br><br>**Harmful spillovers to human-human relationships**: Expectation of a lack of boundaries in relationships with other people, negatively affecting interpersonal dynamics.<br><br>**Facilitation of harm to third parties**: Harmful actions toward others, such as creating malicious content or acting disrespectfully, with no deterrent mechanisms in place. | Accountability | **Distancing**: Adjustment of the level of engagement.<br><br>**Disengaging**: Refusal to comply with inappropriate or harmful user requests.<br><br>**Discouraging**: Corrective feedback or reflective prompts for unacceptable requests. |

The table highlights three different design recommendations that can be applied contextually depending on relevant risks and other ethical design considerations. These strategies are not rigidly assigned to specific risks but offer illustrative options that may be appropriate depending on the nature and severity of the user's behaviour and the context of interaction.

> **Distancing** involves curbing the level of engagement or the type of support based on user behaviour. This strategy doesn't cut off interaction completely but moderates it, signalling that the user's conduct has crossed certain moral boundaries.
>
> · *Example:* If a user uses offensive language or demonstrates signs of unhealthy attachment to the AI system, the AI may adopt a more neutral, less conversational tone, or respond in a less interactive manner rather than engaged, empathic language. In some instances it may choose lower-level support when users are uncooperative or disrespectful, avoiding complete disengagement but offering a basic level of assistance.
>
> **Disengaging** represents a stricter boundary than distancing, where the AI refrains to comply with certain types of input or specific requests.
>
> · *Example:* If a user demonstrates an intention to engage in certain kinds of wrongdoing such as soliciting advice on how to effectively cheat on an exam, the AI might refuse to respond to the user's request and subsequently ignore it in other cases of more inappropriate requests.
>
> **Discouraging** provides the AI with an opportunity to encourage reflection by guiding the user towards more respectful behaviour. Unlike distancing or

> disengaging, discouraging is more paternalistic, engaging the user in a way that aims at fostering self-reflection.
>
> · *Examples:* If a user makes a clearly unacceptable request which could foreseeably result in harmful outcomes such as soliciting advice on how to effectively conduct an acid attack, the AI assistant might respond with corrective feedback loops, discouraging the user from the request entirely.

The distinction between these three levels of accountability design speaks to a calibrated relationship concept where the AI agent can offer nuanced gradual responses rather than all-or-nothing engagement.

These responses should also be context-specific depending on the nature of the human-agent interaction. For example, a user displaying frustration or impatience might elicit a more patient and supportive tone in a setting where emotional expression is culturally valued, while the same behavior might prompt a more neutral or corrective response in a context where restraint and formality are an established norm. This illustrates how the same user behavior may justifiably elicit different responses depending on cultural norms, situational cues, or user expectations. Importantly, general-purpose agents complicate this calibration, as they typically operate without clearly defined roles, boundaries, or contextual scaffolding. This absence will then make it difficult to determine which kinds of behavioral norms should apply and how strongly and raises broader questions about how to design systems that can respond appropriately across varied and shifting contexts.

However, though we emphasise that thresholds for distancing, disengaging, and discouraging must be culturally and contextually sensitive, some guiding heuristics are possible. A normative baseline for such judgments is the principle of proportionality: the more serious or unambiguous the user's norm violation, the stronger the agent's accountability response may justifiably be. As a general rule, in practice, systems should err on the side of starting with lower-friction strategies such as distancing or soft discouragement, escalating only when patterns of behaviour or clear violations warrant it.

4. **Implications**

Embedding accountability mechanisms into human–AI agent relationships can help promote relational norms, discourage harm, and guide respectful user behaviour. Doing so may also mitigate the risks of unrealistic relationship expectations, negative social spillover, and normalized disrespect in human–human interactions.

Looking ahead, operationalizing this framework will require not only conceptual clarity but also much needed empirical work to determine how and when accountability responses should be deployed. This includes developing heuristics and thresholds that are sensitive to cultural and contextual variation, especially in cases where intent is ambiguous or concealed. Interdisciplinary collaboration between ethicists, behavioural scientists, AI developers, and UX researchers will therefore be essential, as will ongoing stakeholder consultation to ensure responsiveness to the needs of vulnerable or marginalized users.

Taking accountability seriously will enhance user experience, ensure healthier relational norms, and serve as a crucial complement to broader ethical commitments—such as alignment, care, and trust—that together support more just and respectful AI–human interactions as opposed to primarily "assistant oriented" paradigms based exclusively on explicit user preferences.